# Stabilization of multi-beam necklace solitons in circular arrays with spatially modulated nonlinearity

Yaroslav V. Kartashov,[1] Boris A. Malomed,[2] Victor A. Vysloukh,[1] and Lluis Torner[1]

[1]ICFO-Institut de Ciencies Fotoniques, and Universitat Politecnica de Catalunya, Mediterranean Technology Park, 08860 Castelldefels (Barcelona), Spain

[2]Department of Physical Electronics, School of Electrical Engineering, Faculty of Engineering, Tel Aviv University, Tel Aviv, 69978, Israel

We address necklace solitons supported by circular waveguide arrays with out-of-phase modulation of nonlinearity and linear refractive index. Such two-dimensional necklace solitons appear as rings of multiple out-of-phase bright spots. We show that necklace solitons are stable if their peak intensity is low or moderate and that the domain of stability shrinks with increasing nonlinearity modulation depth. However, we find that the stability domains expand with increasing number of necklace spots.

The properties of self-trapped optical beams, or solitons, are affected substantially by transverse modulations of the nonlinear material where they propagate. In addition to modulation of linear refractive index, current fabrication technologies allow modulation of nonlinearity, thus geometries with periodic nonlinearity modulation are of particular interest. In contrast to usual lattices that impact beam propagation for any peak intensity, the effects caused by nonlinear lattices become more apparent at high peak intensity, while low-intensity beams remain largely unaffected. Solitons in purely nonlinear lattices have been addressed in both one- and two-dimensional settings [1-7]. An even more interesting situation is realized when linear and nonlinear lattices coexist and compete with each other, e.g. in the case of out-of-phase modulation of focusing nonlinearity and linear refractive index. Such a competition may result in intensity-controlled shape transformations, modifications of soliton center location, enhancement of transverse mobility, and drastic stability modifications [8-18]. All these phenomena were discussed for fundamental solitons, simple multipoles, and vortex solitons.

An interesting class of complex two-dimensional self-trapped states exists in the form of so-called necklace solitons. Such collective states composed out of multiple bright spots arranged into ring-like configurations were first studied in uniform nonlinear media [19-28]. In uniform medium necklaces composed from out-of-phase spots expand upon propagation [19,20], but proper engineering of phases of beams in necklaces allows to construct pulsating and even quasi-stationary complexes. In cubic, saturable, or quadratic medium necklaces are unstable, but they can be metastable in materials with competing quadratic-cubic or cubic-quintic nonlinearities [25,26]. Light bullets may also form necklace solitons [27]. Recently necklace solitons were observed experimentally in local uniform medium [28]. Modulation of linear refractive index of the material may result in stabilization of necklace solitons. In particular, necklace solitons composed even from out-of-phase bright spots may propagate undistorted in the presence of linear lattice [29-33]. Necklace solitons have been studied only in lattices with spatially-homogeneous nonlinearity. In this paper we study their properties in circular waveguide arrays with out-of-phase modulation of nonlinearity and linear refractive index. They bifurcate from linear guided modes of the array and are stable at moderate peak intensities. In contrast to multipole solitons supported by linear lattices, whose stability domain usually shrink with increasing number of poles [29-33], we find that in the geometry addressed here, increasing number of spots in the necklace results in a considerable expansion of the necklace stability domain.

To describe the propagation of light beams along the $\xi$ axis in a circular waveguide array with out-of-phase modulation of linear refractive index and nonlinearity coefficient we use the nonlinear Schrödinger equation for the field amplitude $q$:

$$i\frac{\partial q}{\partial \xi} = -\frac{1}{2}\left(\frac{\partial^2 q}{\partial \eta^2} + \frac{\partial^2 q}{\partial \zeta^2}\right) - [1 - \sigma R(\eta,\zeta)]|q|^2 q - pR(\eta,\zeta)q. \tag{1}$$

Here $\eta, \zeta$ are the transverse coordinates normalized to the characteristic transverse scale $r_0$ (for example, beam width) and $\xi$ is the longitudinal coordinate normalized to the diffraction length $k_0 r_0^2$ corresponding to the selected transverse scale $r_0$, $k_0$ is the wavenumber, $p$ and $\sigma$ are the depths of modulation of linear refractive index and nonlinearity. The function $R(\eta,\zeta) = \sum_{k=1}^{n} \exp[-(\eta-\eta_k)^2/d^2 - (\zeta-\zeta_k)^2/d^2]$ describes an array of $n$ Gaussian waveguides with widths $d = 1/2$ and centers $(\eta_k, \zeta_k)$ located on a ring of radius $nr_{\min}/2$, where a minimal radius $r_{\min} = 0.6$ is achieved for a system of two waveguides at $n = 2$. Note that the radius of the array increases linearly with $n$ so that the length of arc between adjacent waveguides remains constant for any $n$. The nonlinear coefficient $\gamma = 1 - \sigma R$ attains minima at the points where the refractive index has a maximum. Here we set the depth of the linear lattice at $p = 7$ and study the impact of $n$, hence the number of necklace pearls, and of the nonlinearity modulation depth $\sigma$ on the properties of the necklaces. Equation (1) conserves the energy flow $U$ and the Hamiltonian $H$:

$$\begin{aligned} U &= \int\int_{-\infty}^{\infty} |q|^2 \, d\eta d\zeta, \\ H &= \frac{1}{2}\int\int_{-\infty}^{\infty} [|\partial q/\partial \eta|^2 + |\partial q/\partial \zeta|^2 - 2pR|q|^2 - \gamma|q|^4] d\eta d\zeta. \end{aligned} \quad (2)$$

We search for the profiles of necklace solitons in the form $q = w(\eta,\zeta)\exp(ib\xi)$, where $b$ is the propagation constant. Upon linear stability analysis, we substitute perturbed solutions $q = [w + u\exp(\delta\xi) + iv\exp(\delta\xi)]\exp(ib\xi)$ (here $u(\eta,\zeta), v(\eta,\zeta)$ are small perturbations and $\delta = \delta_r + i\delta_i$ is the growth rate) into Eq. (1) and linearize it to get the eigenvalue problem

$$\begin{aligned} \delta u &= -\frac{1}{2}\left(\frac{\partial^2 v}{\partial \eta^2} + \frac{\partial^2 v}{\partial \zeta^2}\right) + bv - \gamma v w^2 - pRv, \\ \delta v &= +\frac{1}{2}\left(\frac{\partial^2 u}{\partial \eta^2} + \frac{\partial^2 u}{\partial \zeta^2}\right) - bu + 3\gamma u w^2 + pRu, \end{aligned} \quad (3)$$

which can be solved numerically. Representative profiles of necklace solitons are shown in Fig. 1. They are composed of $n$ bright spots located in different waveguides of the array with phase changing by $\pi$ between neighboring spots. Such solitons bifurcate from antisymmetric linear guided modes of circular array. At low and moderate $U$ weakly-localized neighboring spots in the necklace strongly interact [Figs. 1(a) and 1(c)]. The positions of intensity maxima are close to centers of Gaussian waveguides, but they are slightly shifted in radial direction due to net radial force arising because of repulsion between nearest neighbors in necklace. Increasing the propagation constant at moderate values results in the gradual contraction of the spots and thus in energy concentration inside each waveguide. For larger $b$ values (corresponding to higher peak intensities) the nonlinear contribution to refractive index becomes dominating, so that light is attracted to the regions of stronger focusing nonlinearity. In this regime necklace soliton transforms into a set of almost noninteracting narrow bright spots strongly shifted in the radial direction [Figs. 1(b) and 1(d)]. Thus, one gets a nonmonotonic $U(b)$ dependence [Fig. 2(a)].

Necklace solitons exist for propagation constant values above a cutoff $b_{\text{co}}$. This cutoff coincides with the propagation constant of the corresponding linear mode and increases monotonically with $p$; it does not depend on the nonlinearity modulation depth $\sigma$ and only slightly changes with growth of $n$. For a fixed $b$, the energy flow of necklace soliton increases with growing $\sigma$. When the propagation constant exceeds a certain value $b_{\text{in}}$, the energy flow drops off when increasing $b$. Thus, high-intensity bright spots focus into the regions of higher nonlinearity. When $b \to \infty$, light is self-trapped in the regions where both nonlinearity and linear refractive index are almost uniform. In this limit the shape of each bright spot is very close to the profile of the Townes soliton of uniform cubic medium. Therefore, the energy flow of necklace soliton asymptotically approaches the value $U = nU_{\text{T}}$, where $U_{\text{T}} = 5.85$. Note that the maximal energy flow carried by the necklace varies little with $\sigma$ [Fig. 2(a)]. The $H(U)$ diagrams exhibit a single cuspidal point corresponding to the propagation constant value $b = b_{\text{in}}$ where $dU/db = 0$ [Fig. 2(b)].

Interestingly, we found that the stability properties of necklace solitons depend strongly on the depth of nonlinearity modulation. Necklaces are stable not only at low energy flows (which is expected, since necklaces bifurcate from linear guided modes), but also at moderate $U$ values, in the fully nonlinear regime (Fig. 3). Necklaces become unstable when peak intensity increases and necklaces undergo strong shape transformations. The instability appears at a propagation constant value $b_{\text{cr}}$ that is smaller than $b_{\text{in}}$ corresponding to the inflection point of the $U(b)$ curve. Typical stability and instability domains on the plane $(\sigma, b)$ are shown in Fig. 3(a) for a necklace with $n = 10$. The width of stability domain shrinks rapidly with growth of $\sigma$. Note that necklace solitons exist and can be stable even for $\sigma > 1$, although it may be difficult to realize such situation in practice, because for such $\sigma$ values the nonlinearity becomes defocusing around waveguide centers.

The central finding of this paper is that at fixed $\sigma$ the width of stability domain $\delta b = b_{\text{cr}} - b_{\text{co}}$ considerably expands for increasingly complex necklace, i.e. for higher-pearl necklaces [Fig. 3(b)]. This is clearly in sharp contrast to multipole solitons in linear lattices, that as expected on intuitive grounds become more sensitive to perturbations with increase of the number of poles [29-33]. We attribute such unexpected stabilization at higher $n$ to the fact that in necklaces with large number of pearls and large radius, each spot interacts effectively only with two adjacent ones (notice that the length of the arc between two waveguides does not vary with $n$). In contrast, in necklaces with small number of pearls, where the radius of necklace is comparable with the arc length, each spot interacts with more neighbors resulting in destabilization at smaller $b$ values. For larger and larger num-ber of pearls the width of the stability domain approaches an asymptotic value. A typical dependence of the perturbation growth rate on $b$ is shown in Fig. 3(c). The slope of such dependence changes abruptly at $b = b_{\rm in}$ where $dU/db$ changes its sign. The growth rate is a complex number, with a non-vanishing imaginary part, for $b_{\rm cr} < b < b_{\rm in}$ (oscillatory-type instability) and becomes purely real for $b > b_{\rm in}$ (exponential instability).

Notice, that properties of necklace solitons depend substantially also on widths of Gaussian waveguides arranged into ring configuration. In particular, increase of the width of waveguides is accompanied by the growth of maximal energy flow carried by the necklace. Thus, for $\sigma = 0.7$ and $n = 6$ the maximal energy increases from $U_{\rm max} \approx 50.3$ for $d = 0.4$ to $U_{\rm max} \approx 68.8$ at $d = 0.7$. At the same time, when $d$ becomes sufficiently small, the overlap between waveguides vanishes and separate spots forming the necklace in high-amplitude limit behave like individual solitons concentrating entirely inside individual waveguides and do not experiencing any radial displacements (this trend was found already at $d \sim 0.2$). While cutoff for soliton existence $b_{\rm co}$ and $b_{\rm in}$ value monotonically increase with growth of waveguide widths $d$, the actual width of stability domain $b_{\rm cr} - b_{\rm co}$ decreases with $d$. Thus, at $\sigma = 0.7$, $n = 6$ one has $\delta b \approx 3.5$ for $d = 0.4$, while for $d = 0.7$ the width of stability domain is given by $\delta b \approx 2.2$. This is because the overlap between waveguides increases with increase of their widths that, in turn, enforces the interactions between neighboring spots in necklace soliton.

Direct simulations of perturbed necklaces confirmed always the results of the linear stability analysis. Stable necklaces propagate undistorted over indefinitely long distances [Figs. 4(a) and 4(b)]. During the propagation of unstable necklaces with $b > b_{\rm in}$, their pearls shift into the regions of almost uniform nonlinearity and undergo collapse. Unstable necklaces with $b_{\rm cr} < b < b_{\rm in}$ decay via progressively increasing oscillations [Figs. 4(c) and 4(d)].

Summarizing, we analyzed necklace solitons in circular waveguide arrays with out-of-phase modulation of nonlinearity and linear refractive index. Such solitons undergo remarkable power-dependent shape transformations. The stability domain of the necklaces shrinks with increasing the nonlinearity modulation depth, but we found that it expands with growing number of necklace pearls.

# Figure captions

Figure 1 (color online). Field modulus distributions for necklace solitons with $n=4$ (a),(b), and $n=10$ (c),(d) for different $b$ values and $\sigma=1$. White rings indicate circular array of waveguides. All quantities are plotted in arbitrary dimensionless units.

Figure 2 (color online). Energy flow versus $b$ (a) and Hamiltonian versus energy flow (b) for different $\sigma$ values at $n=4$. Circles in (a) and (b) correspond to solitons in Figs. 1(a) and 1(b). All quantities are plotted in arbitrary dimensionless units.

Figure 3. (a) Stability (white) and instability (shaded) domains on the $(\sigma,b)$ plane for necklace solitons with $n=10$. (b) $b_{\mathrm{cr}}$ and $b_{\mathrm{in}}$ versus $n$ at $\sigma=0.7$. (c) Real part of perturbation growth rate versus $b$ at $n=4$, $\sigma=0.7$. All quantities are plotted in arbitrary dimensionless units.

Figure 4 (color online). Stable propagation of necklace soliton with $n=10$ corresponding to $b=2.7$ (a),(b) and decay of unstable necklace with $n=10$ corresponding to $b=3.3$ (c),(d). Field modulus distributions are shown at different distances $\xi$. In all cases $\sigma=1$. White rings indicate circular array of waveguides. All quantities are plotted in arbitrary dimensionless units.

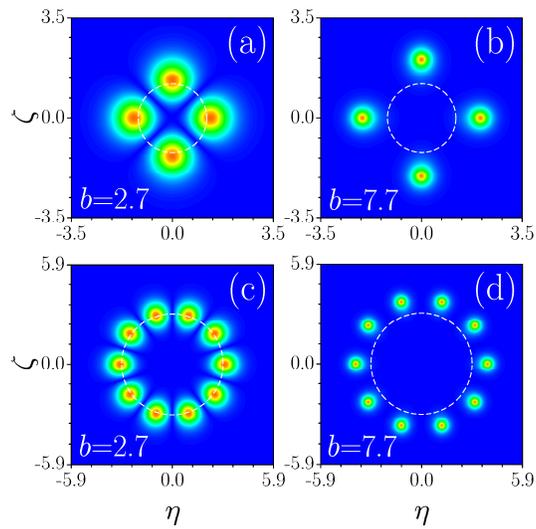

Figure 1 (color online). Field modulus distributions for necklace solitons with $n = 4$ (a),(b), and $n = 10$ (c),(d) for different $b$ values and $\sigma = 1$. White rings indicate circular array of waveguides. All quantities are plotted in arbitrary dimensionless units.

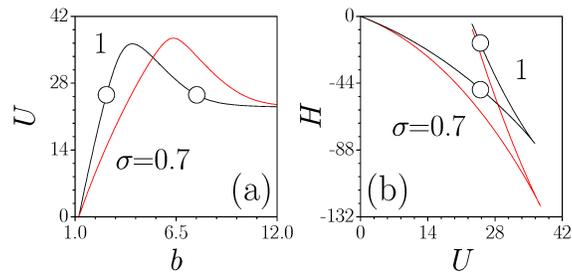

Figure 2 (color online).   Energy flow versus $b$ (a) and Hamiltonian versus energy flow (b) for different $\sigma$ values at $n = 4$. Circles in (a) and (b) correspond to solitons in Figs. 1(a) and 1(b). All quantities are plotted in arbitrary dimensionless units.

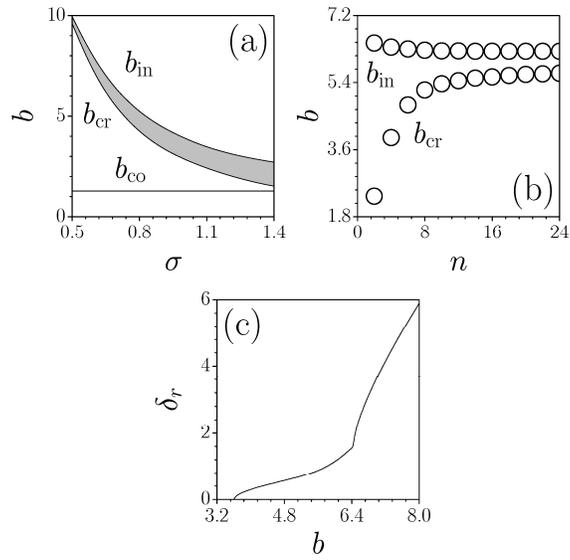

Figure 3. (a) Stability (white) and instability (shaded) domains on the $(\sigma,b)$ plane for necklace solitons with $n=10$. (b) $b_{\mathrm{cr}}$ and $b_{\mathrm{in}}$ versus $n$ at $\sigma=0.7$. (c) Real part of perturbation growth rate versus $b$ at $n=4$, $\sigma=0.7$. All quantities are plotted in arbitrary dimensionless units.

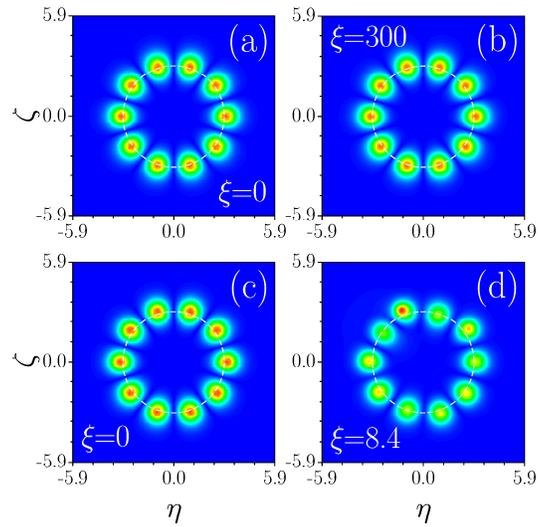

Figure 4 (color online).  Stable propagation of necklace soliton with $n = 10$ corresponding to $b = 2.7$ (a),(b) and decay of unstable necklace with $n = 10$ corresponding to $b = 3.3$ (c),(d). Field modulus distributions are shown at different distances $\xi$. In all cases $\sigma = 1$. White rings indicate circular array of waveguides. All quantities are plotted in arbitrary dimensionless units.

13